# Agricultural Knowledge Management Using Smart Voice Messaging Systems: Combination of Physical and Human Sensors

Naoshi Uchihira[1] and Masami Yoshida[2]

[1] Japan Advanced Institute of Science & Technology, Nomi, Ishikawa 923-1292, Japan
E-mail: uchihira@jaist.ac.jp

[2] Agrisensing LLC, Sapporo, Hokkaido 003-0063, Japan
E-mail: agrisensing@mbr.nifty.com

**Abstract**
The use of the Internet of Things (IoT) in agricultural knowledge management systems is one of the most promising approaches to increasing the efficiency of agriculture. However, the existing physical sensors in agriculture are limited for monitoring various changes in the characteristics of crops and may be expensive for the average farmer. We propose a combination of physical and human sensors (the five human senses). By using their own eyes, ears, noses, tongues, and fingers, farmers could check the various changes in the characteristics and conditions (colors of leaves, diseases, pests, faulty or malfunctioning equipment) of their crops and equipment, verbally describe their observations, and capture the descriptions with audio recording devices, such as smartphones. The voice recordings could be transcribed into text by web servers. The data captured by the physical and human sensors (voice messages) are analyzed by data and text mining to create and improve agricultural knowledge. An agricultural knowledge management system using physical and human sensors encourages to share and transfer knowledge among farmers for the purpose of improving the efficiency and productivity of agriculture. We applied one such agricultural knowledge management system (smart voice messaging system) to a greenhouse vegetable farm in Hokkaido. A qualitative analysis of accumulated voice messages and an interview with the farmer demonstrated the effectiveness of this system. The contributions of this study include a new and practical approach to an "agricultural Internet of Everything (IoE)" and evidence of its effectiveness as a result of our trial experiment at a real vegetable farm.

## 1 INTRODUCTION

In recent years, a combination of the Internet of Things (IoT), cloud computing, and artificial intelligence as a cyber-physical system is producing various innovations and applications [1,2,3]. In this paper, "IoT" in a broad sense refers to this combination of IoT (physical side) and cloud computing with artificial intelligence (cyber side). IoT applications include "smart factories," "smart grids," "smart cities," "smart transportation and logistics," "smart homes," and "smart healthcare." In addition, "smart agriculture" is one of the most promising approaches for increasing the efficiency and productivity of agriculture.

In particular, the rapid aging of farmers has become a serious social problem, especially in Japan. Aged farmers have plenty of tacit knowledge based on their experiences. However, there are insufficient means and efforts for them to externalize and transfer their knowledge to younger generations. Hence, using IoT to record and analyze agricultural data and farming activities is a very important and promising approach to externalizing such tacit knowledge.

Recently, there have been many studies and practical applications of IoT in agriculture. Many researchers focus on physical sensors and networks to visualize agricultural processes [4]. Some have proposed the application of machine-learning techniques to agriculture [5].

However, existing physical sensors, including weather, $CO_2$, and soil sensors, are functionally limited and not widespread enough to monitor various changes in crop characteristics and conditions. They are also expensive for the average farmer. For example, the detection of pests in crops is very difficult for physical sensors. Successful detection is possible but rather expensive to achieve. Therefore, we propose the combination of physical and human sensors, which are the five human senses. Farmers could check the various changes in the characteristics and conditions (colors of leaves, diseases, pests, faulty or malfunctioning equipment) by use of their eyes, ears, noses, tongues, and fingers while verbally recording their observations (Fig. 1). Supporting methods and tools for recording and analyzing such human sensor data would be required.

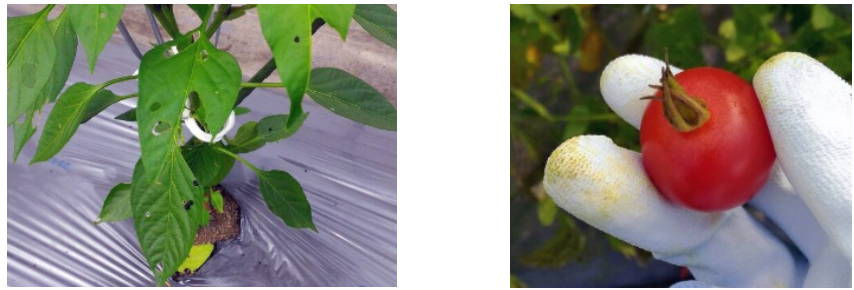

Fig.1 Characteristics and conditions of crops that cannot be detected by traditional physical sensors.

To record observations, we propose a hands-free smart voice messaging system (which functions as an audio version of Twitter) that we have developed for healthcare services [6,7,8] in the JST RISTEX service science program. Since the accuracy of voice recognition has seen recent drastic improvements, voice messages can be transcribed to text data on a server. The data captured by the physical and human sensors (voice messages) are analyzed by data and text mining to improve improving the efficiency and productivity of agriculture in an agricultural knowledge management system.

Our approach can be termed as an "Internet of Everything (IoE)," which includes both the IoT and Internet of Human Sensing (IoH). We have applied this system to a greenhouse vegetable farm in Hokkaido and demonstrated its effectiveness through a qualitative analysis of accumulated voice messages and an interview with the farmer. We describe an agricultural knowledge management system using the smart voice messaging system in Section 2. Then, we discuss the system's effectiveness in Section 3.

## 2 AGRICULTURAL KNOWLEDGE MANAGEMENT USING SMART VOICE MESSAGING SYSTEM

### 2.1 Combined Use of Physical and Human Sensors

We propose the combined use of physical and human sensors in an agricultural IoE. Figure 2 shows the data and knowledge processing flow, which consists of visualizations, pattern extractions, and knowledge externalization, from the human sensors.

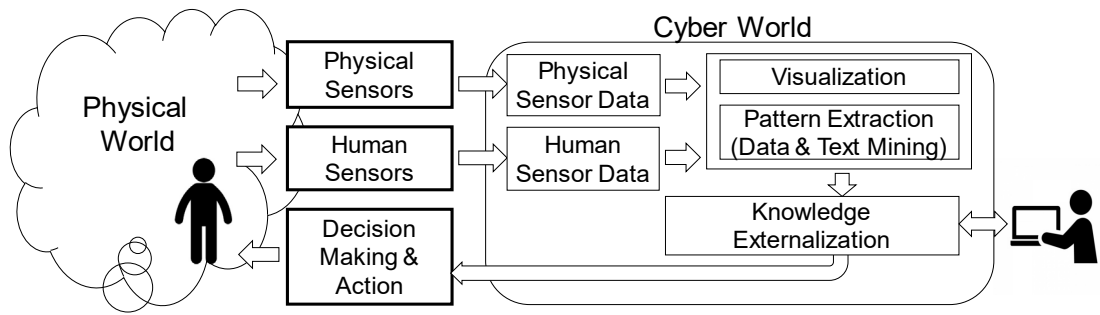

Fig.2 Combined Use of Physical & Human Sensors

Visualization

Raw physical sensor data (mainly numerical data) and human sensor data (mainly text data) can be visualized in a more understandable form. Daily, weekly, and monthly summary reports are useful for understanding trends of observed and controlled objects. The combined use of physical and human sensor data brings two types of benefits: (1) The user can understand the reasons for the recording of specific physical data (e.g., abnormal data) by confirming with the corresponding text data (e.g., human observations and messages). (2) The user can understand the environmental conditions in which the human observations were made by looking at the physical sensor data.

Pattern Extraction

Some patterns can be extracted by data mining, text mining, and integrated data and text mining. Integrated approaches have been investigated and implemented [9,10].

Knowledge Externalization

The user can externalize the knowledge by interpreting the visualized data and extracted patterns. Externalized knowledge is used for decision-making and taking action. There are two types of knowledge externalization: (1) The user finds new knowledge based on visualized data and extracted patterns. (2) The user confirms his tacit knowledge recalled by visualized data and extracted patterns.

## 2.2 Smart Voice Messaging System

Smart voice messaging system is a hands-free voice recording, distribution, and visualization tool for physical and adaptive intelligent services [6,7,8]. The system has three functions: (1) The user records his voice messages with related information (user ID, time, and location). (2) The system stores these messages (voice and recognized text) in a database and automatically distributes them to other users who are relevant to the messages. (3) The system visualizes messages stored in the database from several viewpoints (user, time, location, keyword). Figure 3 shows the outline of the system architecture, which consists of a smartphone application (Java program on Android OS) and a web server application (PHP program on Windows Server). The voice interface of the smartphone application enables hands-free messaging with a headset device. The user's location is recorded outdoors by the GPS function of the smartphone and indoors by Bluetooth low-energy (BLE) beacons. A voice recognition engine (Toshiba RECAIUS) is used as a Platform as a Service (PaaS). The user can interactively use the visualization functions with the web browser and smartphone application. This system can output all or selected information from the database to a CSV file. Integrating data and text mining is possible by using various tools for the CSV file combined with physical sensor data.

This smart voice messaging system is a general-purpose platform and has been used for several physical and adaptive intelligent services, including nursing and caregiving service, equipment inspection and maintenance service, customer service, event security service, and smart agriculture.

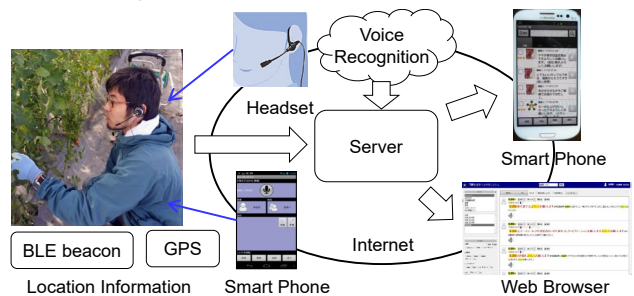

Fig.3 Smart Voice Messaging System

## 2.3 Agricultural Knowledge Management System

Figure 4 shows the proposed agricultural knowledge management system, which uses the smart voice messaging system. Physical sensors include temperature, humidity, CO2, solar radiation, and soil sensors. Human sensors include the records of agricultural activities and the awareness memos and messages sent during these activities. Farmers make daily records using the physical and human sensor data in which reflection and planning are considered. This system supports visualizations, pattern extractions, and knowledge externalization, as explained in Section 2.1. Externalized knowledge is used not only for decision-making in agricultural activities but also knowledge transfer from senior to junior farmers with less experience.

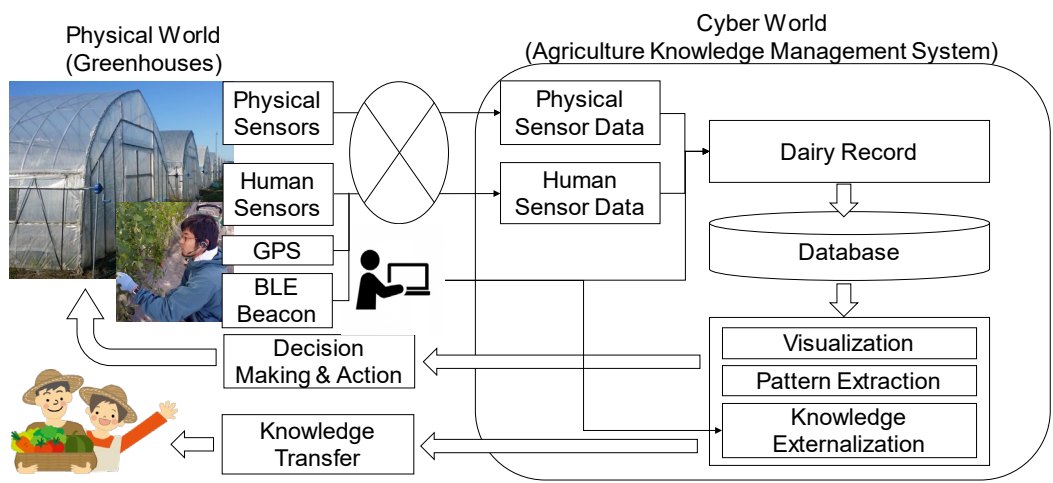

Fig.4 Agricultural Knowledge Management System

## 3 TRIAL EVALUATION

### 3.1 Objectives of Trial Evaluation

The proposed agricultural knowledge management system is still under research and development. To evaluate the effectiveness of the smart voice messaging system, we applied it to a real agricultural project as a trial run. The objectives of the trial included: (1) What kind of voice messages would a farmer record? (2) What kind of voice messages would be effective for knowledge management? (3) What would be the possibilities for the combined use of physical and human sensors?

### 3.2 Evaluation Field

A trial evaluation of the system was conducted at a greenhouse horticultural farm named "Jiyu Kojo" located in Kuriyama, Hokkaido. Vegetables, including green peppers and tomatoes, were being cultivated in 10 greenhouses. The trial evaluation consisted of two terms (June–July and September–November 2017). The owner of this farm had prior experience working in an IT company as a system engineer and was already accustomed to using computers and smartphones. During the trial run, the owner recorded 214 voice messages, of which 200 were valid. After the trial run, we interviewed the farmer on the type, importance, informativeness, and background conditions of each voice message for four hours on 28 January 2018.

### 3.3 Message Classification

Table 1 shows the voice messages classified by subject (farm products, equipment, etc.). More than half of the total number of messages were about farm products (mainly related to pest control) and equipment (mainly related to cleanliness and malfunctions).

Table1 Message Classification by Subject

|   | Subject |    |
|---|---|---|
| 1 | Farm Products | 67 |
| 2 | Equipment | 60 |
| 3 | Sales and Marketing | 24 |
| 4 | Environment (Weather, etc.) | 8 |
| 5 | System (Smart Voice Messaging System, etc.) | 23 |
| 6 | Others | 18 |
|   | Total | 200 |

Table 2 shows a classification by the importance of the messages, i.e., if the message had been worth recording and contained content that was useful, and if so, was worth considering. At least one-fourth of the messages were important and useful.

Table2 Message Classification by Importance

|   | Message Importance |    |
|---|---|---|
| 1 | No importance/not necessary to record | 26 |
| 2 | Importance that can not be judged/necessary to record | 10 |
| 3 | Low importance/useful to record | 89 |
| 4 | Medium importance/useful to record & consider | 41 |
| 5 | High importance/mandatory to record & consider | 12 |
| - | Unclassified (System Problem, etc.) | 24 |
|   | Total | 202* |

*Two messages ware divided into four messages for classification.

Table 3 shows a classification based on the message type and level of informativeness. There are three types: record, action, and consideration. Informativeness of the messages indicates how much more informative than traditional daily records. Although more than half of the information had been recorded in the traditional way (A0, B0), the farmer mentioned that the numbers of A0 and B0 may increase with the smart voice messaging system that can prevent omissions. A more important point is the disappearance of some of the information without this system (A1, A2, B1, B2, C1, and C2). In particular, messages classified into real-time considerations of observations and actions (C2) were worthy of attention. These messages are discussed in Section 3.7.

Table3 Message Classification by Type and Informativeness

|    | Message Type and Informativeness |       |
|----|----------------------------------|-------|
| A0 | Record/no additional information | 75    |
| A1 | Record/quantitative additional information | 5 |
| A2 | Record/ qualitative additional information | 7 |
| B0 | Action/ no additional information | 27   |
| B1 | Action/ quantitative additional information | 6 |
| B2 | Action/ qualitative additional information | 7 |
| C1 | Consideration/quantitative additional information | 4 |
| C2 | Consideration/ qualitative additional information | 24 |
| -  | Unclassified (System Problem, etc.) | 50 |
|    | Total                            | 205** |

**Five messages ware divided into ten messages for classification.

### 3.4 Examples of Important Messages

We present three important messages (Level 5 in Table 2) classified into A2, B2, and C2 in Table 3.

<u>Type A2 message:</u>

<<Voice Message>>

*"Tuscany Violet. Powdery mildew can be seen in the young leaves at the bottom. Although it may have been suppressed by previous mildew control, I will remove these leaves at the time of the next harvest."*

<<Farmer's comment in interview>>

*"This message was worth recording and was not simply about discovering powdery mildew but also the subtle nuance of the relationships among powdery mildew, location and mildew/pest control."*

A description of the relationship among diseases, the environment, and various controls was very important to record because such a description leads to knowledge externalization.

<u>Type B2 message:</u>

<<Voice Message>>

*"In the fourth green pepper house (House #4), the promotion of root growth in the eighth south side of the second row is necessary because the roots are slightly weak. First of all, it is necessary to consider using liquid or pinpoint fertilizer. As the growth of the south side looks to be bad as a whole, we should consider whether this is a problem of water or fertilizer. It may be necessary to spray on the leaves."*

<<Farmer's comment in interview>>

*"The quality of the information is high because there is a description of a specific location." "Perhaps, I used more liquid fertilizer at this time. I thought that the amount of moisture was insufficient."*

Using the information about a specific location, the farmer took appropriate action. Detailed information had been recorded by using real-time and on-the-spot voice messaging.

<u>Type C2 message:</u>

<<Voice Message>>

*"Careful treatment is required this season since the green pepper can be easily ripped from its root. If this happens, the product value will become zero. Please be careful."*

<<Farmer's comment in interview>>

*"This message is for others." "I noticed this phenomenon last year for the first time but did not understand its cause. Usually, it is not so easy to rip away from the root unless someone applies strength to do so."*

Know-how was shared among users of the system. This message was a confirmation of long-standing concerns.

### 3.5 Effectiveness of Human Sensors (Early Pest Detection)

The early detection of pests (mildew, leaf spots, aphids, armyworms, etc.) is very difficult for traditional physical sensors and is one of the most important functions of human sensors. In this trial evaluation, 18 of 200 voice messages were related to the detection of pests and to pest control. Most of the messages can be ranked as important and useful. This farm tried to avoid pesticides as much as possible. The farmer mentioned that it was very important for someone to detect pests as early as possible in order to minimize the use of pesticides.

### 3.6 Combination of Physical and Human Sensors

There are two types of combined use of physical and human sensor data. The farmer can understand the environmental conditions by using the physical sensor data when the observations (awareness) were recorded as messages. We give two examples:

<<Voice Message>>

*"The sunlight comes and the CO2 concentration in the greenhouse falls sharply from the XXXs to the YYYs. This means that the photosynthetic power is up but the outside is cold and the wind is strong, so it is difficult at this time to decide whether to open the side window for ventilation or not."*

Figure 5 shows the changes in the CO2 sensor data on the day the farmer recorded this message.

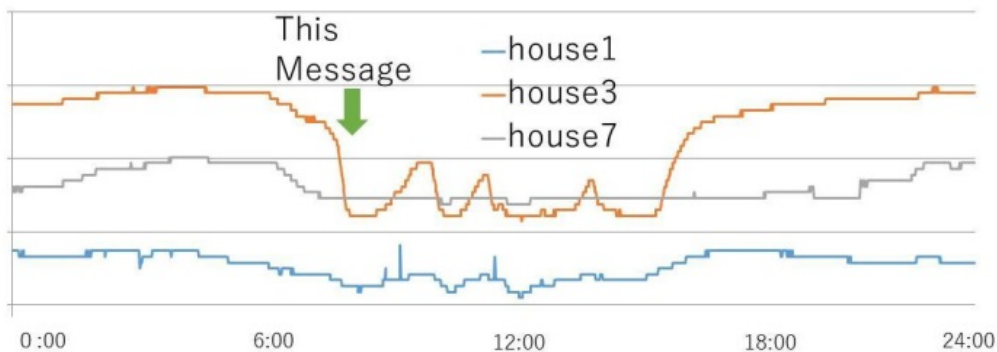

Fig.5 Changes in CO2 Sensor Data

<<Voice Message>>

*"About XX tomatoes near the entrance of House #A are frozen, so House #A needs wind protection. If we do not take any action, the same situation will happen next year."*

Figure 6 shows the changes in the temperature sensor data on the day the farmer recorded this message.

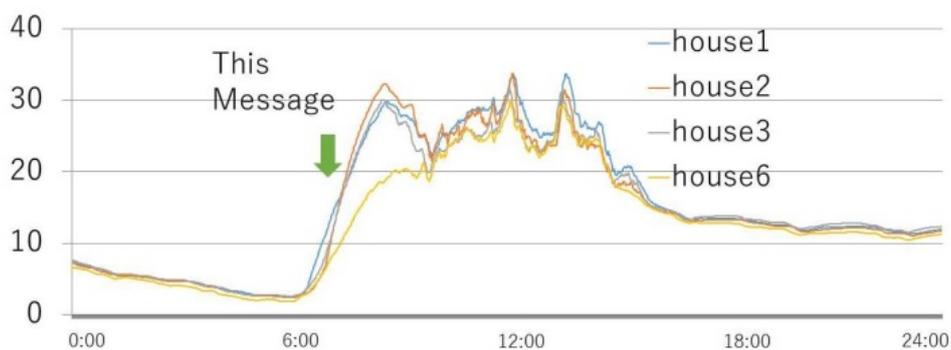

Fig.6 Changes in Temperature Sensor Data

These physical data may be useful for knowledge externalization. Another type of combined use (data analysis using data and text mining) requires more data to be recorded over several years. Therefore, the effectiveness of data and text mining could not be evaluated in this trial run.

### 3.7 Real-time Reflections

The farmer recorded many messages that contained his reflections on his own observations and actions (C1, C2). This is a new way of using the smart voice messaging system. In the interview, he mentioned that he wanted to externalize and share his thoughts on the issues, troubles, and trial-and-error processes he encountered. He expected that the feedback and responses from other users could suggest solutions. This real-time reflection is a kind of the reflection-in-action proposed by Schön [11]. These reflections-in-action using the smart voice messaging system could be expected to help the farmer improve his professional skills.

## 4　CONCLUSION

We have proposed an agricultural knowledge management system utilizing a smart voice messaging system with human sensors and evaluated the potential effectiveness of the human sensors (voice messages) in a trial run at a greenhouse horticultural farm in Hokkaido. We confirmed that some of the voice messages were more informative than the traditional way of making records (diary record). An interesting finding was the farmer's real-time reflections on his own observations (awareness) and actions in some of the informative messages.

One of the limitations of this study is the trial evaluation's having been conducted over only one season and with only one farmer. Pattern extraction by data and text mining, as well as knowledge externalization, requires more data collected over several seasons and with several farmers. The establishment of voice message templates in an agricultural domain is one of the most important directions for future research.

### ACKNOWLEDGMENTS

We would like to thank Mr. Hitoshi Horita of "Jiyu Kojo" for his great contribution to the trial evaluation of the smart voice messaging system. We also thank Mr. Takashi Kinoshita for his assistance in the trial evaluation. This research is supported by Grant-in-Aid for Scientific Research of MEXT under Grant No. 15H02785.

### REFERENCE

[1] Holler, J., Tsiatsis, V., Mulligan, C., Avesand, S., Karnouskos, S., Boyle, D., 2014, From Machine-to-machine to the Internet of Things: Introduction to a New Age of Intelligence, Academic Press, Inc.

[2] Da Xu, L., He, W., Li, S., 2014, Internet of things in industries: A survey. IEEE Transactions on industrial informatics, 10(4), 2233-2243.

[3] Whitmore, A., Agarwal, A., Da Xu, L., 2015, The Internet of Things - A survey of topics and trends. Information Systems Frontiers, 17(2), 261-274.

[4] Liqiang, Z., Shouyi, Y., Leibo, L., Zhen, Z., Shaojun, W., 2011, A crop monitoring system based on wireless sensor network. Procedia Environmental Sciences, 11, 558-565.

[5] McQueen, R. J., Garner, S. R., Nevill-Manning, C. G., Witten, I. H., 1995, Applying machine learning to agricultural data. Computers and electronics in agriculture, 12(4), 275-293.

[6] Uchihira, N., Torii, K., Chino, T., Hiraishi, K., Choe, S., Hirabayashi, Y., Sugihara, T., 2016, Temporal-Spatial Collaboration Support for Nursing and Caregiving Services, in Global Perspectives on Service Science: Japan (Eds. James C. Spohrer, Stephen K. Kwan, Yuriko Sawatani), Springer: 193-206.

[7] Uchihira, N., Choe, S., Hiraishi, K., Torii, K., Chino, T., Hirabayashi, Y., Sugihara, T., 2013. Collaboration Management by Smart Voice Messaging for Physical and Adaptive Intelligent Services, PICMET2013: 251-258.

[8] Torii, K., Uchihira, N., Hirabayashi, Y., Chino, T., Yamamoto, T., Tsuru, S., 2016, Improvement of Sharing of Observations and Awareness in Nursing and Caregiving by Voice Tweets, in Serviceology for Designing the Future (Eds. Takashi Maeno, Yuriko Sawatani, Tatsunori Hara), Springer: 161-175.

[9] Drewes, B., 2002, Integration of text and data mining. WIT Transactions on Information and Communication Technologies, 28.

[10] SPSS, PASW Text Analytics, http://www.spss.com.hk/software/modeling/text-analytics/whats_new.htm  (accessed Aug 14, 2018).

[11] Schön, D. A., 1984, The reflective practitioner: How professionals think in action. Basic Books.